\documentclass[10pt]{sigchi}
\makeatletter
\def\@copyrightspace{\relax}
\makeatother

\usepackage{balance}       
\usepackage{graphics}      
\usepackage[T1]{fontenc}   
\usepackage{txfonts}
\usepackage{mathptmx}
\usepackage[pdflang={en-US},pdftex]{hyperref}
\usepackage{color}
\usepackage{booktabs}
\usepackage{textcomp}

\usepackage{microtype}        

\usepackage{ccicons}          

\def\plaintitle{SIGCHI Conference Proceedings Format}

\def\emptyauthor{}
\def\plainkeywords{Authors' choice; of terms; separated; by
  semicolons; include commas, within terms only; this section is required.}

\makeatletter
\def\url@leostyle{%
  \@ifundefined{selectfont}{
    \def\UrlFont{\sf}
  }{
    \def\UrlFont{\small\bf\ttfamily}
  }}
\makeatother
\def\pprw{8.5in}
\def\pprh{11in}

\definecolor{linkColor}{RGB}{6,125,233}
\hypersetup{%
  pdftitle={\plaintitle},
  pdfauthor={\emptyauthor},
  pdfkeywords={\plainkeywords},
  pdfdisplaydoctitle=true, 
  bookmarksnumbered,
  pdfstartview={FitH},
  colorlinks,
  citecolor=black,
  filecolor=black,
  linkcolor=black,
  urlcolor=linkColor,
  breaklinks=true,
  hypertexnames=false
}

\newcommand{\dhat}{\operatorname{\hat{d}}}
\newcommand{\mhat}{\operatorname{\hat{m}}}
\newcommand{\vecx}{\operatorname{\boldsymbol{x}}}
\newcommand{\Complex}{\mathbb{C}}

\begin{document}

\title{End-to-End Radio Fingerprinting with Neural Networks}

\numberofauthors{5}
\author{%
  \alignauthor{Ryan M. Dreifuerst\\
    \email{ryandry1st@utexas.edu}}\\
  \alignauthor{Andrew Graff\\
    \email{andrewgraff@utexas.edu}}\\
  \alignauthor{Sidharth Kumar\\
    \email{sidharth.kumar@utexas.edu}}\\
  \alignauthor{Clive Unger\\
    \email{clive.unger@utexas.edu}}\\
  \alignauthor{Dylan Bray\\
    \email{dfbray@utexas.edu}}\\
}

\maketitle

\begin{abstract}
    
    This paper presents a novel method for classifying radio frequency (RF) devices from their transmission signals. Given a collection of signals from identical devices, we accurately classify both the distance of the transmission and the specific device identity. We develop three models, each building off the previous, that lead to a multiple classifier system that accurately discriminates between channels and classifies devices using normalized in-phase and quadrature (IQ) samples from the data. The first two are a device-classifier and distance-classifier, with the third combining the two for an end-to-end ensemble classifier. Our residual network (ResNet) based architecture reaches $88.33\%$ accuracy classifying 16 unique devices over 11 different distances and two different runs, on a task that was previously unlearnable. Furthermore, we demonstrate the efficacy for pre-training neural networks that need to identify subtle features from massive data domains.

\end{abstract}



\section{Introduction}

With the advent of 5G and other wireless standards, an increasing number of devices will be connected to the internet to assist various vertical applications, such as autonomous driving, telemedicine, industrial applications, and drone communications. As the number of these low cost IoT devices increases, security from illicit eavesdroppers is a primary concern. To mitigate malicious eavesdroppers, this work looks into the problem of distinguishing RF devices from their transmission fingerprints, commonly known as as RF fingerprinting. Along with security improvements, new applications like device identification in heterogeneous networks, vehicle recognition by self-driving cars and network regulation could be improved by RF fingerprinting. 

Every transmitting device has subtle inherent differences, even following the same transmission protocol, which are reflected in the transmitted signals. 
Through radio fingerprinting, a receiver tries to learn these differences by characterizing devices by distinguishable features like its power amplifier non-linearities. 
This process is further complicated by the dynamic nature of wireless channels, which affects both the power and the phase of the incoming signals.
In this work, we train a neural network classifier to identify devices by directly learning the differences from the IQ samples of extremely precise X310 USRP software-defined radios (SDRs).

Besides RF fingerprinting, some other aspects of wireless communication where machine learning is applicable are channel estimation and symbol detection \cite{Farsad2018}, frequency estimation \cite{Drei2020}, spectrum sensing (cognitive radio) \cite{zigbeeCNN}, resource allocation using multi-armed bandit \cite{Gus2019}, energy modelling \cite{AZM2016}, and end-to-end communication systems \cite{Felix2018}. 

Given a set of signals from bit-similar, identical devices, our goal is to accurately predict both the distance of each device from the receiver and to identify which specific device the signal came from, by capturing the unique hardware irregularities of each device. This paper presents 3 models, each building on the previous, that lead to an end-to-end multiple classifier system that accurately discriminates between channels and classifies devices using normalized in-phase and quadrature samples from the dataset presented in \cite{Sankhe2020}. These 3 models are a device-classifier, a distance-classifier, and an ensemble-classifier that combines the first two into an end to end network. We found that our ResNet based architecture is able to reach over $88.33\%$ accuracy in classifying devices over 11 different channels and two different runs, on a task that was previously not successfully learnable.

\subsection{Prior work}
In the past, researchers have looked at many different methods to learn device characteristics, but we narrow our focus to data driven algorithms for radio fingerprinting.
\cite{Sankhe2020} proposed a convolutional neural network (CNN) called "ORACLE", where they intentionally introduced impairments at the transmitter to increase accuracy, hence suppressing the effects of wireless channels. 
The work in \cite{IoT2018}, explores this problem by identifying four different features, i.e. frequency offset, I/Q offset, modulation offset and differential constellation trace figure. Data was collected for 54 different Zigbee devices and a K-means algorithm is used to cluster the devices.
\cite{zigbeeCNN} also used Zigbee devices and applied a CNN, which used the complex base-band signals of 7 different Zigbee devices for cognitive radio applications, resulting in a reported accuracy of $92.29\%$. 
Kose et. al \cite{Access2019}, proposed a probabilistic neural network classifier trained on the energy spectrum of the transmitter. Specifically, they employed principal component analysis to select the three most important transient signals and train the network on those to reduce computational complexity. 
In many ways these prior works have all revolved around making the task learnable by identifying critical information in the received signal, but none of them consider the effects that the wireless channel has on their classifiers. Thus, our work is twofold: we propose a more powerful network which outperforms prior work, and expand the network to include discrete channel identification to overcome the channel effects. First, we provide a brief description of the dataset, then we define the system model the classifiers will learn and an associated training process, and finally, we demonstrate our algorithms effectiveness with simulation results.

\section{Dataset}
    We use the 16 device test bed dataset from \cite{Sankhe2020}. 
    To collect this dataset, 16 different bit-similar X310 USRP SDRs are placed at a specific distance of \{2, 8, 14, 20, 26, 32, 38, 44, 50, 56, 62\} feet from an X210 USRP SDR which serves as the receiver. The transmitting device then transmits OFDM modulated random data at 5 MS/s with a carrier frequency of 2.45 GHz. The raw IQ samples are then recorded prior to channel estimation or synchronization. In this setup, each of the 16 devices are run twice at each distance for about 0.4 seconds, reaching over $7 \times 10^{9}$ total samples, where each sample is a complex value, $Z[n] \in \Complex$. 
    
    In \cite{Sankhe2020}, the authors chose a sliding window approach to partition the training samples into overlapping windows of length 128. The step size between each window is never specified, however, one graphic suggests they chose a step size of 1, which would enhance the shift in-variance but cause windows to be highly correlated with each other. Alternatively, we divide each I/Q sequence into non-overlapping windows of length 256 to reduce correlation between windows.
    
\begin{figure}[!t]
    	\centering
    	\begin{tabular}{cc}
    	\hspace*{-4mm}\includegraphics[width=0.26\textwidth]{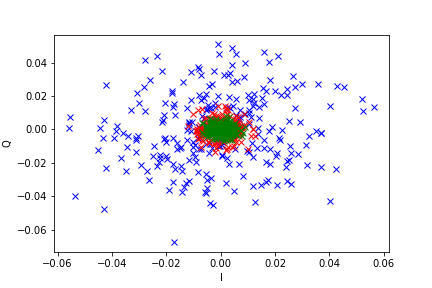} & \hspace*{-6mm} \includegraphics[width=0.26\textwidth]{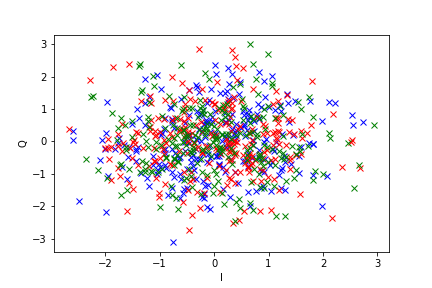}\\
    	Raw & Normalized\\
    	\end{tabular}
    	\caption{One set of raw (left) and normalized data (right) for each distance of 2, 8, and 14 ft.}
    	\label{fig:norm_comp}
    \end{figure}

    We used normalization on each window before training. This served two purposes. First, normalized data generally leads to faster convergence during training. Second, it requires our distance-classifiers to predict distance irrespective of the power of the received data. This second point is particularly compelling, because we force our models to learn differences within the changing phase rather than simply the amplitude of the data. This means that there are complex, non-amplitude changes in the channel that the model can detect and associate with a change in distance, even without knowledge of the symbols being transmitted. In practice, there may be no prior knowledge or consistency to the power that a device decides to transmit at, meaning that the receive power may not be a robust indicator of distance. In Fig.~\ref{fig:norm_comp}, we visualized sets from 3 distances before and after normalization. Before normalization, the amplitude decreases drastically with distance. As such, classifying these sets by distance is trivial for a neural network. After normalization, there is no perceivable amplitude difference between sets. This makes classification by distance much more interesting.
        
\section{System model}
    We begin by defining the set of transmission distances as $D$, and the set of transmitting machines as $M$. Then we propose three models, the first is a function which takes a sequence of IQ samples, $\{\vecx_j[i]\}_0^{N-1}$ and predicts which machine; $\mhat_j$, transmitted the signal assuming the device was at a distance $d = j$. In this setup, the model assumes the machines are all transmitting from the same distance. Then, the goal is to approximate the function $\phi_j(\cdot)$ such that 
    \begin{equation} \label{eqn: mod1}
        \mhat_j = \phi_j(\vecx_j[0], \vecx_j[1] ... \vecx_j[N-1]), \quad \mhat_j \in \{M\},\; j \in \{D\}.
    \end{equation}

    We train separate models for each distance with this learning model in mind, because the differences in channel make the device identification task unlearnable when the channel is inconsistent.
    
    The second model predicts the distance between the transceivers. In this setup, the sequence is the same, but the data is drawn from a set where all of the devices at a specific distance have the same label and normalized samples, but the labels now correspond to the distance that the transmitter was transmitting from. This way, the model learns to recognize channels based on latent channel features such as scattering, fading, and frequency selectivity. The task is made more difficult by the use of random data transmission, reducing the structure and information of the problem such that direct channel state information is unrecoverable. 
    We identify each distance as a separate label and produce an estimate of the distance as 
    \begin{equation} \label{eqn: mod2}
        \dhat = \psi(\vecx_d[0], \vecx_d[1] ... \vecx_d[N-1]), \quad  d, \dhat \in \{D\}.
    \end{equation}
    This model allows us to detect the distance, and is an orthogonal goal to (\ref{eqn: mod1}).
    
    In the final model, we consider the end-to-end system, where we first identify the distance, then classify the device given the selected distance. This is a fundamental problem, as the task of identifying bit-similar devices from random IQ data without CSI has, to the best of our knowledge, never been done successfully. We show that the task of end-to-end learning is possible by combining (\ref{eqn: mod1}) and (\ref{eqn: mod2}) to produce a jointly learnable end to end model of the form
    \begin{align} \label{eqn: e2e}
        \dhat &= \psi(\vecx_d[0], \vecx_d[1] ... \vecx_d[N-1]), \nonumber \\
        \mhat_{\dhat} &= \phi_{\dhat}(\vecx_d[0], \vecx_d[1] ... \vecx_d[N-1]).
    \end{align}
    
    Given the learning model, we now review our contributions which lead to classifiers for (\ref{eqn: mod1})-(\ref{eqn: e2e}).
\section{Our contributions} 

    Given a set of IQ samples, $\{\vecx_d[i]\}_0^{N-1}$, from identical wireless devices our goal is to accurately predict both the distance of the transmitter and the identity of that specific device.
    We do this by modeling the unique hardware irregularities of each device. Several deep learning architectures were developed and tested to achieve this task. We first recreated the simple architecture from \cite{Sankhe2020} but with different window sizes and a more diverse dataset. We then explored the application of a ResNet architecture \cite{He2016} to predict distance and device. Finally we settled on an ensemble of residual networks to first predict the transmission distance then the unique device ID. Each model is trained with various splits of the training data to achieve the network that best generalizes. 

    \subsection{Architectures}

        Initially, we recreated the architecture described in \cite{Sankhe2020} for comparison. We then used a residual convolutional neural network, with residual layers based on Fig. 2 of \cite{He2016}. This model is made of a convolutional layer with 64 filters, then two residual layers with 128 and 256 filters respectively, max pooling, batch-normalization, and three dense layers of 256, 64, and n units, where n is either 11 or 16, corresponding to the cardinality of the distances and devices, $|D|,\ |M|$. Between all convolutional and residual layers there is also max pooling to improve the depth of the feature mapping, as well as element-wise dropout of 0.2 between the fully connected layers to improve generalizability. In all cases the activation function is ReLu, the kernel size is 5, and the inputs are zero-padded to maintain dimensionality. 
        
        We trained these architectures according to (\ref{eqn: mod1})-(\ref{eqn: mod2}) to create 2 types of models: device-classifiers and distance-classifiers. The device-classifiers were trained on subsets of the dataset containing only one distance with the labels corresponding to unique devices. The models would then learn to classify the devices, assuming a static channel. The distance-classifier was trained on subsets of the dataset containing all distances but with the labels corresponding to the distance the transmitter was placed from the receiver. These would then predict which distance a device was transmitting at, irrespective of the particular device.
        
    	\begin{figure}[!ht]
    		\centering
    		\includegraphics[width=0.4\textwidth]{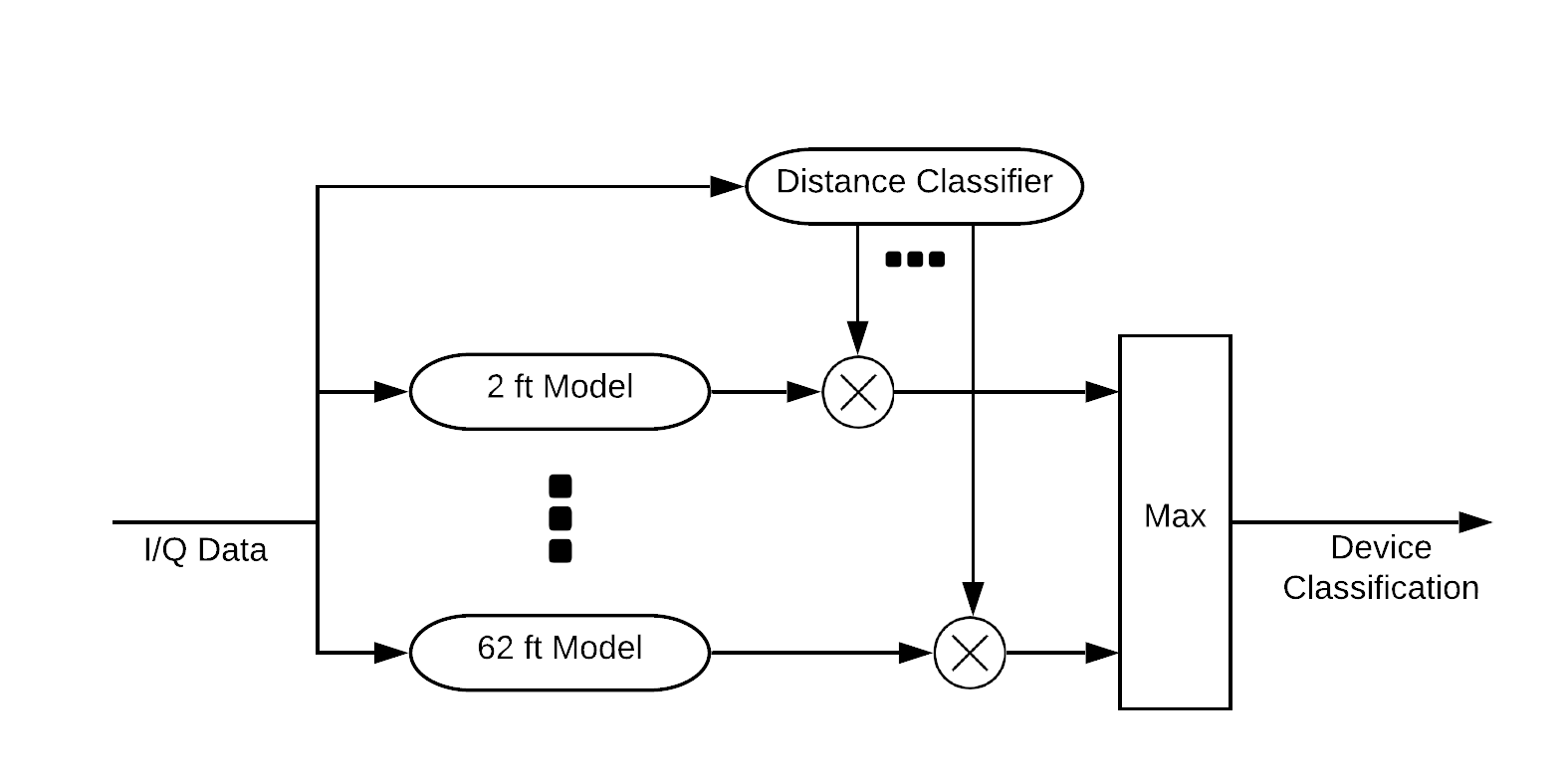}
    		\caption{The ensemble architecture used to classify devices at all distances.}
    		\label{fig:ensemble}
    	\end{figure}
        
        To combine the classification accuracy of each single-distance model, we introduce an ensemble architecture. Our model evaluates the device predictions from all 11 single-distance models and concatenates their one-hot classification outputs. The distance classifier model then creates a binary mask, which is applied to the concatenated outputs, isolating the prediction for the network trained on the best estimated distance. This architecture is visualized in Fig.~\ref{fig:ensemble}.
    
    \subsection{Training Process}
        
        All models were trained using a lab PC with a Nvidia RTX 2080 Ti or on Google Colab cloud based GPUs. Early models trained on all data for a certain distance (i.e. 2 feet) failed because they would not converge or could not perform better than random guessing due to the massive data domain space and subtle differences between devices. We found that by training on a small portion of the data first, then moving to a larger set of data allowed the networks to converge. Each model was first trained on a subset of the original data which had 160,000 samples of 256 length windows. The pre-trained weights were then trained again on a larger dataset of 1,250,400 samples of 256 length windows from the same dataset. Training was performed using gradient descent and the loss defined by the cross-entropy.         To prevent over-fitting, early stopping and learning rate decay are used based on the validation accuracy. 
        
        We partition the train and validation/test data in two ways. One is to group both signals runs together and then split the data. The second method separates the two runs, using one as training and the other as validation/test. Best results were achieved by training on both runs. We suspect the time variance of the channel causes the two runs to be only mildly comparable, making the classifier run-dependent if it only was trained on one set.

    \subsection{Complete System}
        
        The final ensemble model was trained by loading a model trained on each distance, then using the distance classifier referenced earlier to select which of those models to use. Initially, the device prediction model is naively combined with the distance classifier and not re-trained in the context of the ensemble. However, we also consider the case where we retrain the device model to fine-tune the weights.

\section{Simulation results}
\begin{figure}[!ht]
		\centering
		\includegraphics[width=0.4\textwidth]{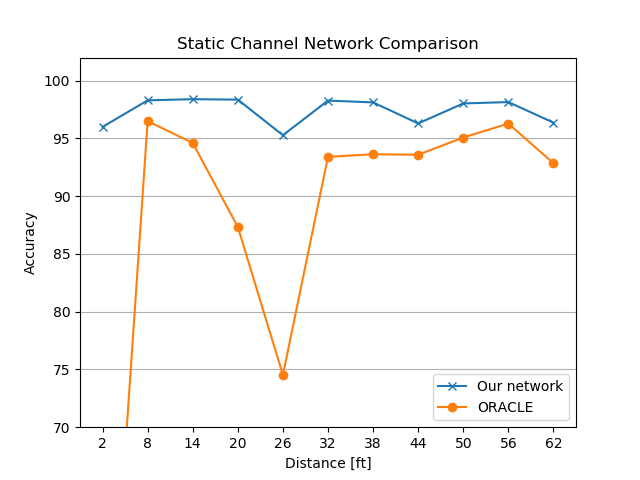}
		\caption{Accuracy results of our model and ORACLE for each distance i.e., static channel. Our results show that the signal to noise ratio is not affecting the accuracy, but that some distances were noticeably more difficult. ORACLE is not able to learn at all at 2ft, likely due to the lack of expressive power of their model combined with receiver saturation, while our networks maintain $95\%$ accuracy or higher.}
		\label{fig:sim1}
	\end{figure}

    We show the results from our models trained on the three tasks from (\ref{eqn: mod1}) - (\ref{eqn: e2e}). 
    We first evaluate our model and compare with the model from  \cite{Sankhe2020}, named "ORACLE", for each distance with normalization. 
    The accuracy for each distance is higher than ORACLE as shown in Fig. \ref{fig:sim1}. These results show that we are able to improve on the original network by training deeper network by preventing vanishing gradients through the residual connections and batch normalization.

	For the learning problem from (\ref{eqn: mod2}), there is no comparable network, as \cite{Sankhe2020} only concluded that the difference in channels made the results inconsistent, likely suggesting that they could not actually learn to identify the devices if the channel was not identical in training and testing. Our distance network was able to reach greater than $98\%$ accuracy for distance prediction on unseen data from all of the devices, suggesting that the channel is learnable from random data, even after normalization. 
	
\begin{figure}[!ht]
		\centering
		\includegraphics[width=0.48\textwidth]{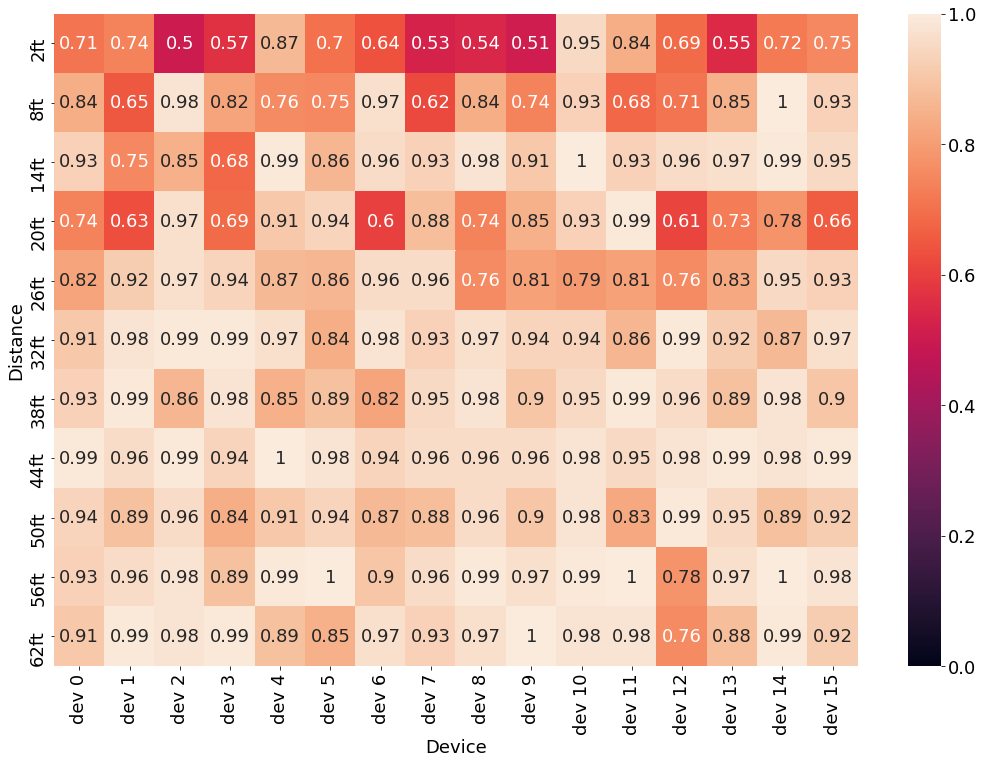}
		\caption{The heat map for our joint distance and transmitter classification algorithm for each distance and device precision. Our model is able to reach very high precision for further distances, while performing poorly for the 2ft, 8ft, and 20ft distances across most devices.}
		\label{fig:sim2}
	\end{figure}

	Using the results from the two models, we test our ensemble network first without any fine-tuning, which resulted in accuracy of more than $95\%$ on data from the same run, but less than $25\%$ accuracy on data from the untrained second run data. We then fine-tune the network using data from each machine, distance, and run, leading to an overall accuracy above $88.33\%$ and the corresponding classification precision map is shown in Fig. \ref{fig:sim2}. 
	
	We can see that receiver saturation is likely a problem, since transmitters very close to the receiver perform poorly. Interestingly, device 10 had generally good performance throughout the test, with an average precision of $94.73\%$. This device may have actually been performing outside of specifications, making it distinctly different from the others. We also note that our network performs best at 44ft, with an average precision of $97.19\%$.
	
    Due to the poor classification accuracy of our ensemble network on second run data prior to fine tuning, it is likely that the features or channel representations learned are time-variant, even though the runs are collected only a few minutes apart. Despite the large size of this dataset, we are limited in that only two runs of data were collected. A more robust system may be learnable if the dataset included several more runs, encompassing the features that change as a result of time. This may, however, be an unlearnable task when allowing for a continuous channel instead of discrete channel identification. We leave this investigation to future work.

\section{Conclusion}
    We have presented an end to end neural network which is able to accurately differentiate channels and classify devices from normalized in-phase and quadrature samples. We found that our ResNet based architecture is able to reach over $88.33\%$ accuracy in classifying devices over 11 training channels and two separate time windows, on a task that  was previously unsuccessful. Furthermore, we have demonstrated the need for pre-training in our networks which are identifying subtle, latent features from massive data domains.
    
    Our study focused only on classification in discrete distances, which is a coarse approximation of true wireless channels. In future work, we will consider how the network can handle continuous, time-varying channels. Additionally, we envision restructuring this problem into a semi-supervised problem where new devices could be added or removed from the network based on clustering techniques.

\bibliographystyle{SIGCHI-Reference-Format}
\bibliography{abbreviations, references}

\end{document}